# Supercontinuum generation in high-index doped silica photonic integrated circuits under diverse pumping settings

C. Khallouf[1], V. T. Hoang[2], G. Fanjoux[1], B. Little[3], S. T. Chu[4], D. J. Moss[5], R. Morandotti[6],

J. M. Dudley[1], B. Wetzel[2], and T. Sylvestre[1]

[1]Institut FEMTO-ST, CNRS-Université de Franche-Comté, 25030 Besançon, France

[2]XLIM Research Institute, CNRS UMR 7252, Université de Limoges, 87060 Limoges, France

[3]QXP Technologies Inc., Xi'an, China

[4]Department of Physics, City University of Hong Kong, Tat Chee Avenue, Hong Kong, China [5]Optical Sciences Centre, Swinburne University of Technology, Hawthorn, Victoria, Australia [6]INRS-EMT, 1650 Boulevard Lionel-Boulet, Varennes, J3X 1S2, Québec, Canada

## ABSTRACT

Recent advances in supercontinuum light generation have been remarkable, particularly in the context of highly nonlinear photonic integrated waveguides. In this study, we thoroughly investigate supercontinuum (SC) generation in high-index doped silica glass integrated waveguides, exploring various femtosecond pumping wavelengths and input polarization states. We demonstrate broadband SC generation spanning from 700 nm to 2400 nm when pumping within the anomalous dispersion regime at 1200 nm, 1300 nm, and 1550 nm. In contrast, pumping within the normal dispersion regime at 1000 nm results in narrower SC spectra, primarily due to coherent nonlinear effects such as self-phase modulation and optical wave breaking. Additionally, we examine the impact of TE/TM polarization modes on SC generation, shedding light on the polarization-dependent characteristics of the broadening process. Moreover, Raman scattering measurements reveal the emergence of two new peaks at 48.8 THz and 75.1 THz in the Raman gain curve. Our experimental results are supported by numerical simulations

---

Further authors information: TS or BW

E-mails: thibaut.sylvestre@univ-fcomte.fr, benjamin.wetzel@unilim.fr

based on a generalized nonlinear Schrödinger equation that incorporates the new Raman gain contribution. Finally, relative intensity noise measurements conducted using the dispersive Fourier transform technique indicate excellent stability of the generated SC spectra.

## 1. INTRODUCTION

Due to their unique combination of high brightness, multi-octave bandwidth, and fiber delivery, supercontinuum (SC) laser sources have revolutionized many applications, including optical coherence tomography (OCT), flu- orescence imaging, optical sensing, absorption spectroscopy, and optical frequency comb metrology.[1–9] While SC light has been extensively researched and developed in optical fibers for over three decades, achieving record bandwidths spanning the ultraviolet (UV) to mid-infrared (MIR) wavelength ranges, current efforts have shifted towards developing compact, low-threshold SC sources based on photonic integrated circuits (PICs). Recent technological advances in the design and fabrication of low-loss highly nonlinear integrated photonic chip waveg- uides are currently driving new research on SC generation in various materials, and yet unexplored regions of the electromagnetic spectrum such as the mid-infrared. SC generation has already been investigated in many integrated and dispersion-engineered photonic integrated circuits based on a range of nonlinear materials, glasses or wide band-gap semiconductors, including silica ($SiO_2$),[10] doped silica (SiON),[11] silicon (Si),[12] silicon nitride ($Si_3N_4$),[13–15] silicon germanium (SiGe),[16,17] titanium dioxide ($TiO_2$),[18] chalcogenide glass ($As_2S_3$),[19] lithium niobate on insulator (LNOI),[20] Aluminum nitride (AiN),[21] and more recently Gallium arsenide (GaAs).[22] For a comprehensive review of the recent progress on chip-based SC sources, we can refer to Reference.[23] Among available nonlinear materials, highly-doped silica glass (HDSG) is quite promising as it possesses several advan- tages including both CMOS and silica fiber compatibility, a wide transparency window from 0.3 μm to 3 μm, a nonlinearity greater than pure fused silica, no free carrier or nonlinear absorption.[11,24] Although SC generation in a such waveguide has already been investigated in the past, the role of Raman scattering, various dispersion pumping regimes, and polarization effects on SC generation hitherto remains elusive.

In this paper, we present new results on SC generation in both short 3 cm and long 50 cm spiral highly- doped silica glass (HDSG) integrated waveguides under various pumping wavelengths and input polarization states. When the waveguides were pumped in the anomalous dispersion (AD) regime (at 1200 nm, 1300 nm, and 1550 nm), broadband SC spectra were generated, spanning from 700 nm to 2400 nm, primarily due to soliton self-frequency shift (SSFS) and dispersive wave (DW) generation. Narrower SC spectra were obtained when pumping the waveguide in the normal dispersion (ND) regime (at 1000 nm), resulting from self-phase modulation (SPM), optical wave breaking (OWB), and also DW generation in the AD regime. Additionally, we investigate the impact of TE/TM pump polarization modes on SC generation. We also report the observation of two new Raman peaks at 48.8 THz and 75.1 THz, in addition to the known fused silica peak at 13.2 THz, using a confocal Raman micro-spectrometer at 532 nm. Using this new Raman gain curve, we found good agreement with numerical simulations based on the modified nonlinear Schrödinger equation. Finally, we present relative intensity noise (RIN) measurements based on the real-time dispersive Fourier transform (DFT) technique.[25]

## 2. HIGH-INDEX DOPED SILICA CHIP WAVEGUIDE

Since the early developments of high-index, low-loss, CMOS-compatible doped silica glass for applications to all-optical signal processing, significant progress has been made, achieving large-scale low-loss (< 0.1 dB/cm) in highly-doped silica glass waveguides. Figure 1(a) shows the cross-section of the doped silica waveguide manufactured by QXP Technologies Inc. The core size is 1.5 $\mu m$ × 1.52 $\mu m$, integrated into plasma-enhanced chemical vapor deposition (PECVD) silica ($SiO_2$), and deposited on $SiO_2$ thermal oxide, all on a silicon wafer. The chip core has a high refractive index $n = 1.7$ at 1550 nm. The chip comprises 20 reversible input/output waveguides, couplers, and delay lines, each connected to 25 cm-long fiber pigtails, facilitating the coupling and extraction of guided light. In our experiments, we investigated SC generation in a 50-cm-long HDSG waveguide, which includes two 25-cm-long spirals. We also compared with the results obtained in a shorter 3-cm long waveguide. The total insertion loss of the long (short) waveguide was measured to be 8.9 dB (3 dB) at 1550 nm using a high-resolution optical time domain reflectometer (OBR 4600 Luna Tech.). The OBR trace of the 50 cm-long waveguide is shown in Fig. 1(b) with a sampling resolution of 20 $\mu m$, and the linear loss was estimated to be as low as 0.1 dB/cm.[26]

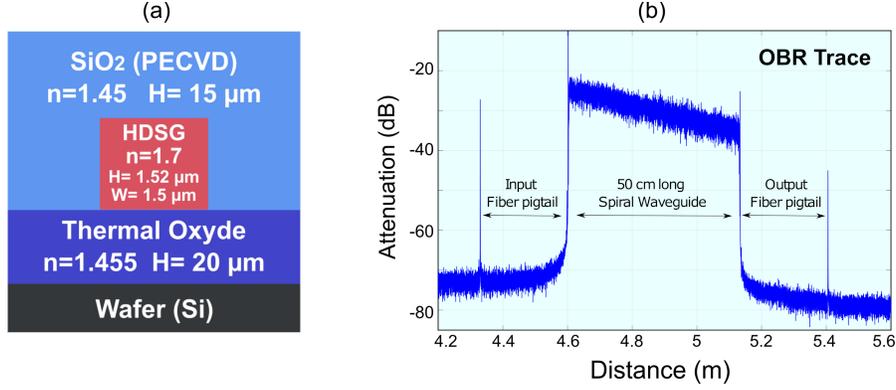

Figure 1: (a) Cross-section of the highly doped silica glass waveguide (HDSG) from QXP technologies; (b): Attenuation trace showing the losses of the 50-cm long waveguide and the fiber pigtails at λ=1550 nm.

## 3. DISPERSION AND EFFECTIVE MODAL AREA

Based on the cross-section of the integrated waveguide shown in Fig. 1(a), we numerically calculated the effective refractive indices of the TE/TM fundamental modes, as well as the various dispersion and non-linearity param- eters. A finite element method (FEM) based software (Ansys Lumerical mode solutions) was used to determine these parameters by scanning the wavelength from 700 nm to 3000 nm, the region of interest for SC generation. The results are provided in Fig. 2, which shows the group-velocity dispersion $D$ in panel (a) and the effective mode area $A_{eff}$ in panel (b) for the two TE/TM fundamental modes, represented in blue and red, respectively. The dispersion profiles for both TE/TM modes exhibit three zero-dispersion wavelengths (ZDW) near 1000 nm, 1800 nm, and 2300 nm. The birefringence of the waveguide was estimated to be below $10^{-4}$ at 1550 nm.

According to Fig. 2, pumping wavelengths will cover both dispersion regimes from 1000 nm to 1550 nm. So when pumping at 1000 nm, i.e, in the normal dispersion regime, one can expect to get SC generation mediated by SPM and OWB, when pumping at 1200 nm, 1300 nm, 1550 nm, i.e in the anomalous dispersion regime, one can expect to have a broader SC generated by high-order soliton fission, Raman-induced soliton self-frequency shift (SSFS), and dispersive wave (DW) emission. Figure 2(b) also demonstrates that the effective modal area $A_{eff}$ of the two fundamental TE and TM modes strongly varies across wavelengths. This variation will impact the efficiency of SC generation and will be considered in the simulations.

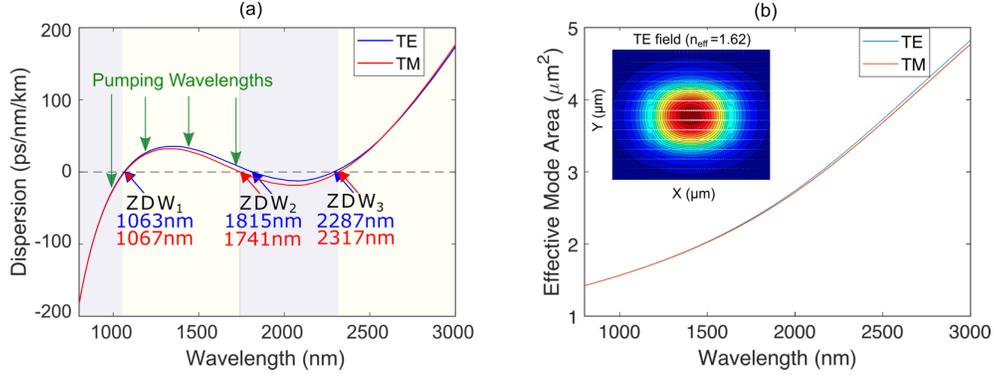

Figure 2: (a): Group velocity dispersion $D$ in ps/nm/km of the QXP-chip waveguide shown in Fig. 1 of the fundamental TE and TM modes in blue and red, respectively, ND/AD is Normal/Anomalous Dispersion; Blue and red arrows indicate the three zero-dispersion wavelengths (ZDW) for each mode, respectively. Green arrows depict the pumping wavelengths considered experimentally. (b): Effective modal area $A_{eff}$ of the two fundamental TE and TM modes versus wavelength. The inset shows the TE optical field transverse distribution.

## 4. RAMAN SPECTRUM OF THE HIGH-INDEX DOPED SILICA GLASS

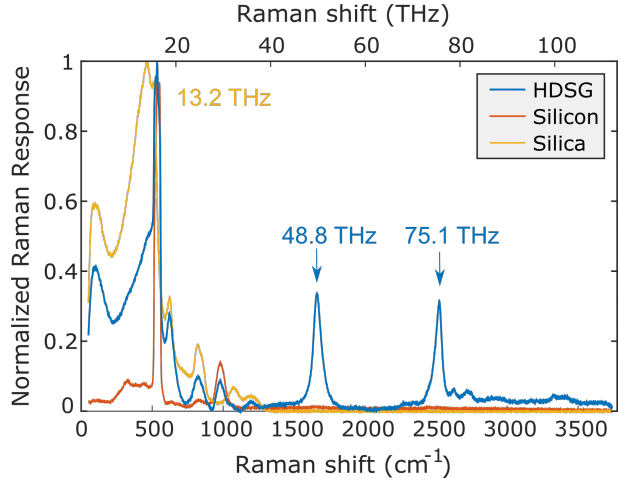

Figure 3: Raman spectra of the integrated HDSG waveguide in blue, of the silicon in orange, and silica in yellow. A normalization was employed for clarity purposes in the presented data.

Before experimentally exploring SC generation, we first examined the Raman response of the high-index material using a high-resolution confocal Raman micro-spectrometer (Monovista CRS+). The Raman spectra presented in Fig. 3 were acquired by scanning from the top of the QXP chip waveguide, employing a 532 nm continuous doubled Nd:YAG laser beam and a 100x microscope objective. Achieving optimal depth focus was essential to ensure precise targeting of the laser beam onto the highly doped silica waveguide, rather than onto the surrounding silica coatings. The Raman spectra of silica (yellow) and silicon (orange) shown in Fig. 3 were obtained by focusing the laser beam onto the upper layer of the PECVD $SiO_2$ layer and the silicon wafer, respectively. During the focusing process on the doped silica material, Raman scattering from silica originating from the surrounding coatings, as well as from the material itself, along with signals from the silicon wafer, were consistently observed. However, notable is the emergence of two distinct peaks at 1627 $cm^{-1}$ (48.8 THz) and 2504 $cm^{-1}$ (75.1 THz), in addition to the usual silica peak at 450 $cm^{-1}$ (13.2 THz). These two new Raman peaks originate from $O_2$ and $N_2$ molecular vibrational states, respectively.[27] This refined Raman gain profile, measured experimentally (in blue), will be further included in our simulations to improve the modeling of SC generation.

## 5. EXPERIMENTAL SETUP

The experimental setup used for generating and characterizing SC spectra is shown in Fig. 4. A 200-fs Optical Parametric Oscillator (OPO) served as the pump laser. It is driven by a Ti-Sapphire mode-locked laser (Chameleon Ultra II) at an 80 MHz repetition rate tunable from 680 to 1080 nm. The OPO signal is tunable within the range 1-1.6 $\mu$m with mean output powers from 230 mW to 1 W, and the idler from 1.7 $\mu$m to 4 $\mu$m with powers ranging from 250 mW to 50 mW. Control over power and polarization angles $\theta$ were achieved through a variable density filter and a half-wave plate, respectively. The OPO signal was then injected into an SMF-28 fiber pigtail with a 55% coupling efficiency using a 40x IR focusing objective mounted on a multi-axis fiber alignment stage. The light was fiber-coupled into the QXP chip waveguides using the input polarization-maintaining fiber pigtail, aligned with the TE/TM axis of the waveguide. The output-generated SC light was analyzed using two different spectrometers to cover the full SC bandwidth. An Agilent 86142B (OSA1) was used for the range 600-1700 nm, with a lower responsivity from 1700 to 2000 nm, and an FTIR Thorlabs (OSA2, model 205C) was used for the range of 1000-5600 nm. To complete the spectral measurements, we also analyzed the real-time SC stability and relative intensity noise (RIN) when pumping at 1550 nm using the dispersive Fourier transform (DFT) technique.[25, 28, 29] This technique utilized a 170-m long dispersion-compensating fiber (DCF) to stretch the pulses and measure the fluctuating pulse-to-pulse SC spectra in the time domain via a dispersive transformation. The output stretched pulses were detected with a high-speed 50-GHz high-power photodiode (FINISAR, model HPDV2120R) and recorded using a 20 GHz bandwidth real-time oscilloscope (Keysight Agilent DSA91204A, 20GHz, 40 GSa/s).[30]

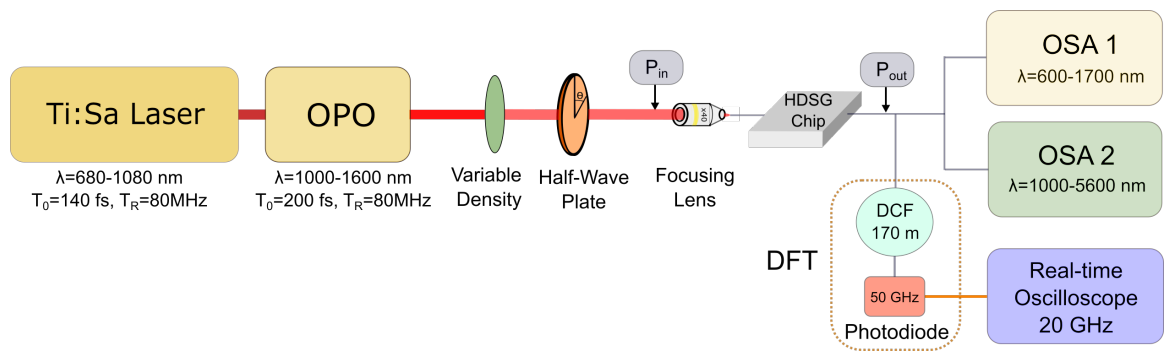

Figure 4: Experimental setup for SC generation and characterization. Ti:Sa Laser: Titanium-Sapphire laser; OPO: Optical parametric oscillator; OSA: Optical spectrum analyzer; DCF: Dispersion compensating fiber. DFT: Dispersive Fourier transform.

## 6. EXPERIMENTAL SC RESULTS

Figure 5 shows the 50 cm long waveguide output SC spectra for three different pumping wavelengths (1550 nm, 1300 nm, 1200 nm) with increasing pump powers for each wavelength setting. As anticipated, the supercontinuum (SC) generated across these various pumping wavelengths arises from soliton self-frequency shift within the AD regime, coupled with dispersive wave generation transitioning into the ND regime. Notably, when employing a pump wavelength of 1550 nm, typical for telecom applications, the resulting SC spectrum spans from 750 nm to well beyond the upper limit of the Agilent analyzer (OSA1) at 2000 nm (Fig. 5(a)), so that extended measurements utilizing the Thorlabs OSA2 were necessary (see Fig. 5(b)). However, to prevent intensity saturation of the Thorlabs spectrometer, the input power at the entrance of the focusing objective was limited to 220 mW, corresponding to an output power of 10 mW. Despite this limited power, a discernible SC is evident, extending up to 2400 nm for a 220 mW input power.

Additionally, Fig. 5(a) illustrates a residual OPO idler located at 1740 nm. However, we verified that this weak idler (- 40 dB compared to the pump) neither contributes nor interferes with the SC generation process. For other pumping wavelengths at 1300 nm and 1200 nm, respectively, idler residues extend beyond 2000 nm and do not appear in the measurements. The SCs depicted in Figs. 5(c) and (d) span from 750 nm to 1800 nm. Here, the input polarization angle was chosen to maximize SC bandwidth, suggesting compatibility with the TE or TM mode. The spectral dip observed at 1380 nm is due to residual water (OH$^-$) absorption which is still present within the integrated waveguide. Nonetheless, this absorption does not impede SC extension towards the infrared.

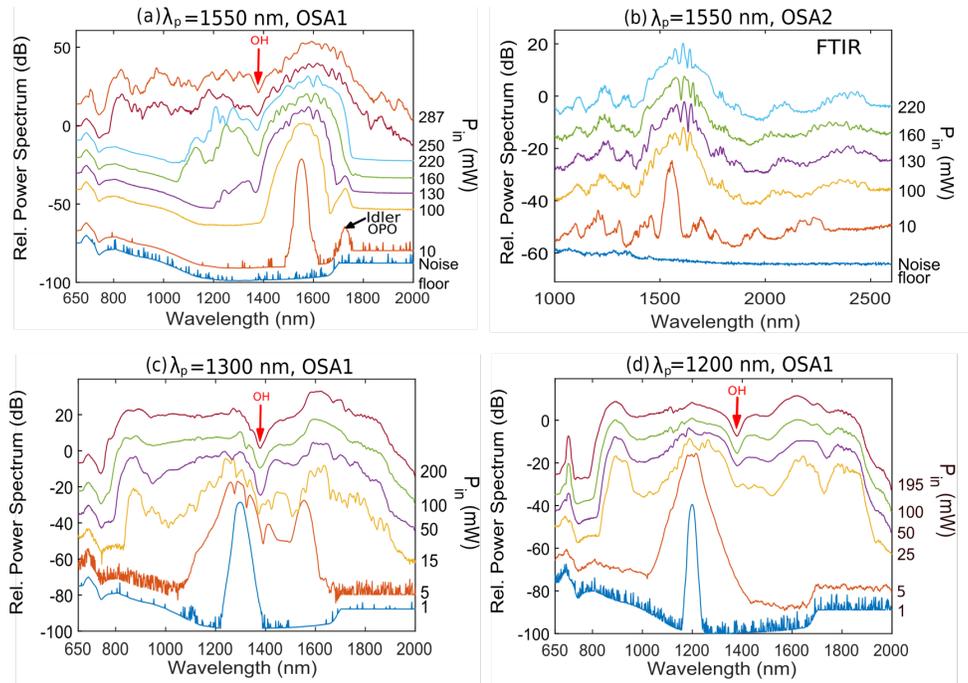

Figure 5: SC spectra generated in the 50 cm waveguide in the anomalous dispersion pumping regime for an increasing input mean power measured at the entry of the focusing objective, shown on the right side of the graph for different pumping wavelengths $\lambda_P$ (1550 nm, 1300 nm, and 1200 nm); (a),(c),(d) were measured with the Agilent OSA1 while (b) was measured with the Thorlabs OSA2. All spectra are vertically shifted for improved clarity.

Figure 6 illustrates spectral broadening obtained when pumping at 1000 nm, within the ND regime and close to the first ZDW. As expected, a limited SC extension is observed, primarily attributed to self-phase modulation (SPM) and optical wave breaking (OWB).[31] At lower power levels, the SC exhibits quasi-symmetric spectral broadening, ceasing upon reaching the first zero-dispersion wavelength (ZDW1) near 1065 nm when pumped at higher powers. Notably, a distinctive peak emerges near 1150 nm, characteristic of dispersive wave generation from OWB within the AD regime, a phenomenon previously observed in optical fibers.[32] This observation will be further corroborated through numerical simulations.

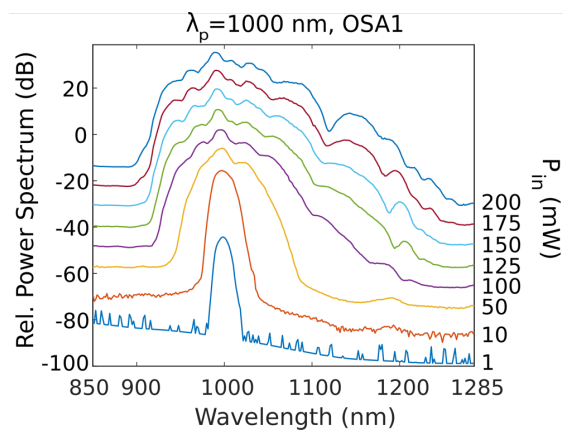

Figure 6: SC spectra generated in the 50 cm waveguide in the normal dispersion regime with increasing input power, shown on the right side of the graph, when pumping at 1000 nm, measured with the Agilent OSA1.

In addition to the experiments reported on the 50 cm spiral waveguide, SC measurements were performed using a 3 cm waveguide having the same material and cross-section. Figure 7 (a) and (b) show the output SC spectra when pumping at 1550 nm and 1300 nm, respectively, with increasing input mean pump power up to 400 mW. Despite the shorter waveguide, the SC still extends from approximately 800 nm to 1900 nm with good flatness. This indicates that the femtosecond SC is mainly generated while propagating in the first few centimeters of the chip waveguide. Consequently, careful design considerations are essential for optimizing SC bandwidth and output power in on-chip SC generation.

## 7. POLARIZATION ANALYSIS

Next, we investigated the SC variations by rotating the input polarization angle relative to the waveguide's TE/TM axes. When pumping at 1550 nm, Fig. 8(a) shows a notable dependency of the SC generation process on the polarization angle. The SC exhibits a wide spectral spread ranging from 800 nm to 2000 nm for specific polarization angles, whereas, for others, it spans over a reduced spectral range from 1150 nm to 1800 nm. This

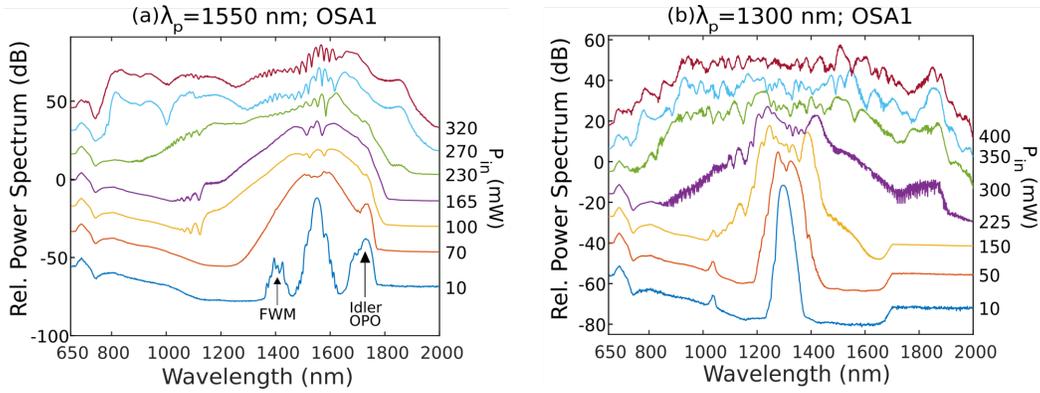

Figure 7: SC spectra generated in the 3 cm waveguide in the anomalous dispersion regime with increasing input power, shown on the right side of the graph when pumping at (a): 1550 nm and (b): 1300 nm measured with the Agilent OSA1.

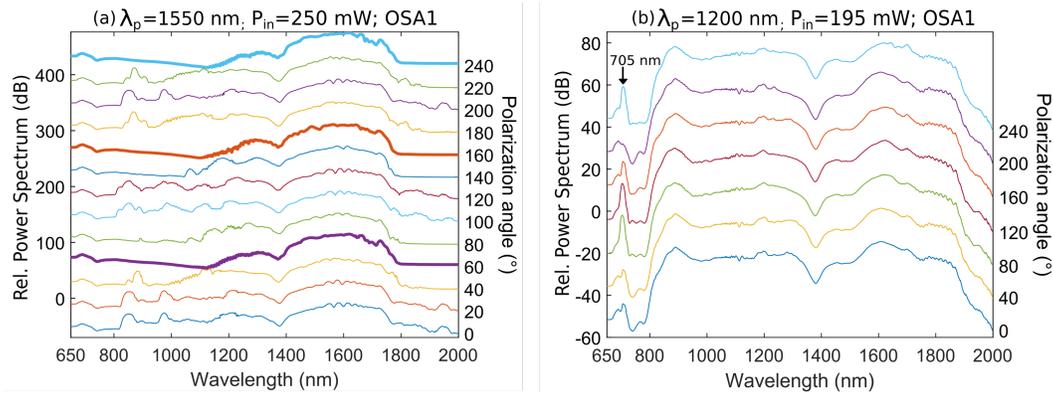

Figure 8: SC generation in the anomalous dispersion regime for different input polarization angles, depicted on the right side of the graph with an input power $P_{in}$ measured at the entry of the focusing objective for a pumping wavelength of (a) 1550 nm with $P_{in}$= 250 mW ; (b) 1200 nm with $P_{in}$= 195 mW. The 90° periodicity of the input polarization angle is illustrated by bold lines for the corresponding SC spectra in (a).

reduced SC efficiency can be easily attributed to pulse broadening due to polarization mode dispersion when the input polarization is not aligned with one of the principal axes of the waveguide. As expected, there is an observed angle periodicity of approximately 90°. For instance, SC spectra exhibit similar characteristics at polarization angles of 60°, 160°, and 240° (see bold lines in Fig. 8(a)). Conversely, when pumping at 1200 nm, as depicted in Fig. 8(b), the polarization angle demonstrates less influence on the SC generated within the 800-2000 nm range. However, a distinct peak near 705 nm displays strong polarization dependence with a periodicity of

180°. This peak has been identified from numerical simulations as resulting from a phase-matched four-wave mixing (FWM) involving the pump at 1200 nm and the dispersive wave at 892 nm. We have checked that the phase-matching condition related to this FWM : $\beta_{892nm} - \beta_{1200nm} - \beta_\lambda = 0$ is satisfied around $\lambda = 705$ nm.

## 8. NUMERICAL SIMULATIONS

To model SC generation within the doped silica chip waveguides, we used the scalar generalized nonlinear Schrödinger equation (GNLSE), including the new Raman gain spectrum and the wavelength-dependent losses. This equation is expressed in the following reduced form[2]

$$\frac{\partial A}{\partial z} + \frac{\alpha(\omega)}{2}A - \sum_{n\geq 2}\frac{i^{n+1}}{n!}\beta_n\frac{\partial^n A(z,T)}{\partial T^n} = i\gamma(1+i\tau_0\frac{\partial}{\partial T})\left(A\int_{-\infty}^{\infty}R(T')|A(z,T-T')|^2 dT'\right), \quad (1)$$

where A(z,T) is the complex amplitude of the field propagating in the *z* direction and in the pump velocity time frame *T*. $\alpha(\omega)$ is the frequency-dependent propagation loss, $\beta_n$ is the $n^{th}$ derivative of the propagation constants. The nonlinear coefficient $\gamma = 2\pi n_2/(\lambda_0 A_{eff}(\lambda_P))$ is obtained from the nonlinear index of the high index silica glass $n_2 = 1.15 \times 10^{-19} m^2/W$ [33] and $A_{eff}(\lambda_P)$, the effective area of the fundamental mode for the pump wavelength $\lambda_P$.[34] The nonlinear coefficient was estimated from the computed modal area plotted in Fig. 2(b), yielding $\gamma = 250\ W^{-1}km^{-1}$ (at 1550 nm), which is approximately 200 times larger than a standard silica fiber (SMF-28). The temporal derivation of the field envelope on the right side of in Eq. (1) corresponds to the self-steepening effect with its characteristic time scale $\tau_0 = 1/\omega_0$ ($\omega_0$ being the input pulse angular frequency). $R(T) = (1 - f_R)\delta(T) + f_R h_R(T)$ is the global nonlinear response, where $h_R(T)$ is the delayed temporal Raman response (assuming the new Raman gain curve shown in Fig. 3), and $f_R$ the fractional contribution of the Raman effect, here considered as $f_R$=0.18, which corresponds to the value of a standard silica fiber.[34] As initial conditions, we consider Gaussian-shaped pulses with 200 fs duration (FWHM), and we also add quantum noise as one photon per mode with random phase as well as the relative intensity noise of the pump laser (1%), as in Ref.[35] In our simulations, we averaged the results of 20 pulse propagations under varying input noise conditions, assuming a polarization aligned with either the fundamental TE or TM mode. Additionally, we take into account the spectral dip resulting from OH⁻ absorption into the simulation as a Gaussian filter centered at 1380 nm.

Finally, the simulations also considered nonlinear propagation in the short fiber pigtails before the integrated waveguide, assuming standard dispersion and nonlinear coefficients for single-mode fibers.

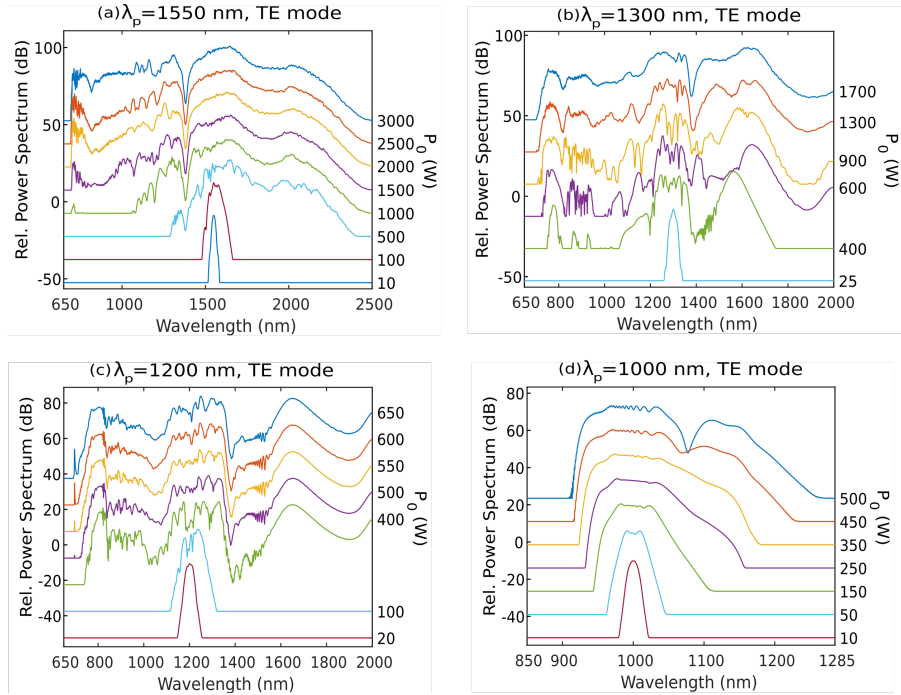

Figure 9: Numerical simulations: SC spectra for different input peak powers $P_0$ shown on the right side of the graph and for different pumping wavelengths $\lambda_P$ as those used experimentally. Simulations are based on the TE mode dispersion.

Figures 9(a-d) present the simulated SC spectra for different input peak powers ($P_0$) while considering the same four pumping wavelengths ($\lambda_P$) used in the experiment. As can be seen, a quite good agreement is observed between these simulations and the experimental data shown in Figs. 5 and 6. Specifically, when pumping at 1550 nm, the simulated SC spreads from 700 nm to 2400 nm (Fig. 9(a)). Similarly, for a pump wavelength of 1300 nm, the simulated SC also extends from 750 nm to 2000 nm (Fig. 9(b)). The peak observed experimentally at 705 nm when pumping at 1200 nm is also replicated in the simulations, with the SC spanning from 700 nm to 2000 nm (Fig. 9(c)). This peak relies on a FWM interaction within the SC, as was discussed earlier.

Furthermore, when pumping at 1000 nm, we observe consistent behaviors between simulations and experi-ments (Fig. 6), particularly the symmetry in the spectra, which breaks upon reaching the first zero-dispersion wavelength (ZDW1) (Fig. 9(d)). An important aspect of the agreement between experiments and TE simulations is that the measurements were predominantly conducted along one of the TE/TM axes. This alignment is justified by the selection of the polarization angle, which aimed to maximize the spectral broadening, noting that SCs generated in the fundamental TM mode exhibited nearly identical characteristics to those in the fundamental TE mode. Nevertheless, a slight difference between experiments and simulations is noticeable, primarily attributed to wavelength-dependent losses (which were not fully characterized except at the telecom wavelength of 1550 nm, i.e. 0.1 dB/cm), but also due to variations in the responses and sensitivities of the two optical spectrum analyzers (OSAs). In terms of pumping power, the mean input powers used experimentally for SC generation agree well with the peak power considered numerically. For instance, when pumping at 1550 nm, a mean power of 200 mW corresponds to a peak power of 2 kW (i.e. pulse energy of 40 nJ), when coupling losses into the waveguide's core are taken into account.

To gain further insight into the physics of SC generation, we have plotted the SC spectro-temporal evolution dynamics in Fig. 10 when considering a pump wavelength of 1550 nm and a peak power of 1.5 kW, which corresponds to the purple spectrum shown in Fig. 9(a). As seen in Figs. 10(c), the pulse is progressively compressed at the onset of propagation so that, after only a few centimeters of propagation, solitons are ejected from the input pulse, and dispersive waves are generated. At this stage (i.e. around 3 cm of propagation), spectral broadening already reached its full extent, as typically observed experimentally in Fig. 7. To analyze dispersive wave (DW) generation, we calculated the phase-matching condition satisfying the equation:
$\beta_{int}$-$\gamma P_S/2$=$\beta(\omega_{DW}) - \beta(\omega_S) - \beta_1(\omega_{DW} - \omega_S)$, with DW and S represents the dispersive wave and soliton, respectively. This result is depicted as a red curve on the left of Fig. 10(b). Phase-matching for the generation of a DW occurs at 710 nm, in good agreement with experimental observations.

## 9. SC COHERENCE AND NOISE

There are two main metrics used to analyze the SC noise properties: the first-order spectral coherence function and the relative intensity noise (RIN). We first modeled the SC coherence using the mutual coherence function,

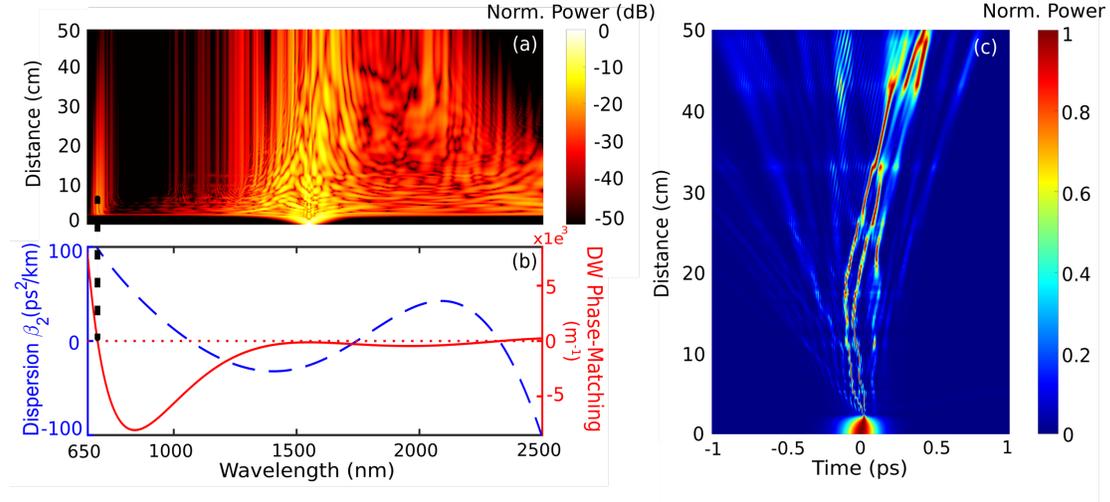

Figure 10: Numerical results: (a): SC spectral broadening dynamics along propagation distance in the 50-cm long integrated waveguide. Numerical simulations are obtained considering a peak power of 1.5 kW at 1550 nm; (b): Dispersion curve (blue, left) and dispersive wave phase matching (red, right) as a function of wavelength. (c): Corresponding temporal evolution dynamics of the SC along propagation in the waveguide.

which can be written in the following form:[2]

$$|g_{12}(\omega)| = \left| \frac{\left\langle \tilde{A}_i^*(\omega)\tilde{A}_j(\omega) \right\rangle_{i \neq j}}{\sqrt{\left\langle \left|\tilde{A}_i(\omega)\right|^2 \right\rangle \left\langle \left|\tilde{A}_j(\omega)\right|^2 \right\rangle}} \right|, \qquad (2)$$

where $[A_i(\omega), A_j(\omega)]$ represent independent pairs of complex spectral amplitude, and angle brackets indicate an ensemble average over independent SC pairs generated from $N$ simulations with different noise seeds. In our simulations, we considered two noise sources including the quantum noise and the relative intensity noise (RIN) of the pump laser (1%), as in Ref.[35] Simulation results are shown in Figs. 11 and 12 for two pump wavelengths at 1000 nm and 1550 nm in the ND and AD regimes, respectively. Specifically, we investigated the broadening and coherence in the TE (Fig. 11) and TM (Fig. 12) modes for different input peak powers. While SC coherence is well preserved in the ND pumping regime at 1000 nm for both TE/TM modes, as shown in Figs. 11(a) and 12(a), coherence is rapidly lost when pumping in the AD regime even at low peak power. This coherence loss is here mainly due to noise amplification processes such as modulation instability and Raman scattering. However, Fig. 12(b) shows that the coherence is fairly well preserved when pumping on the TM mode.

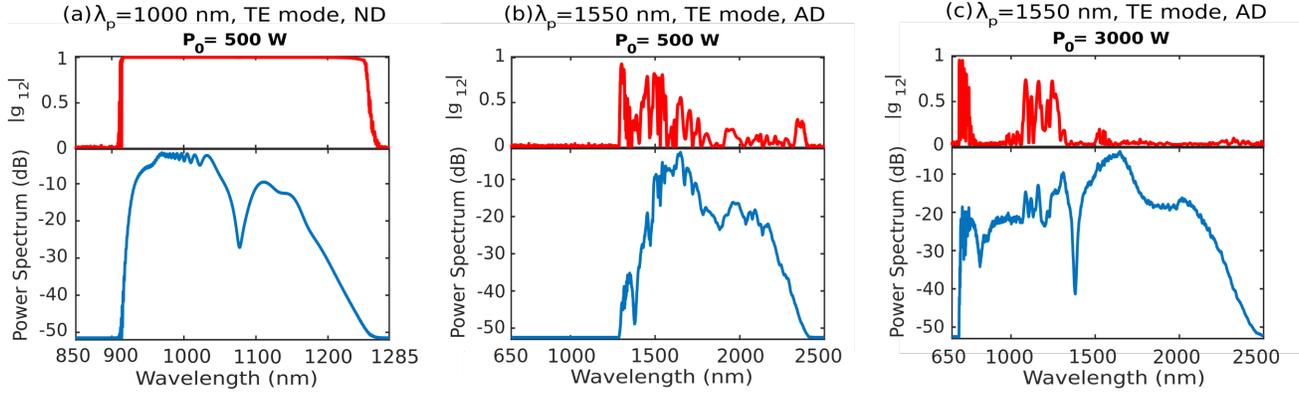

Figure 11: Numerical simulations of SC generation (blue) and mutual coherence (red) at 1000 nm and 1550 nm, respectively, for the TE mode and different peak powers.

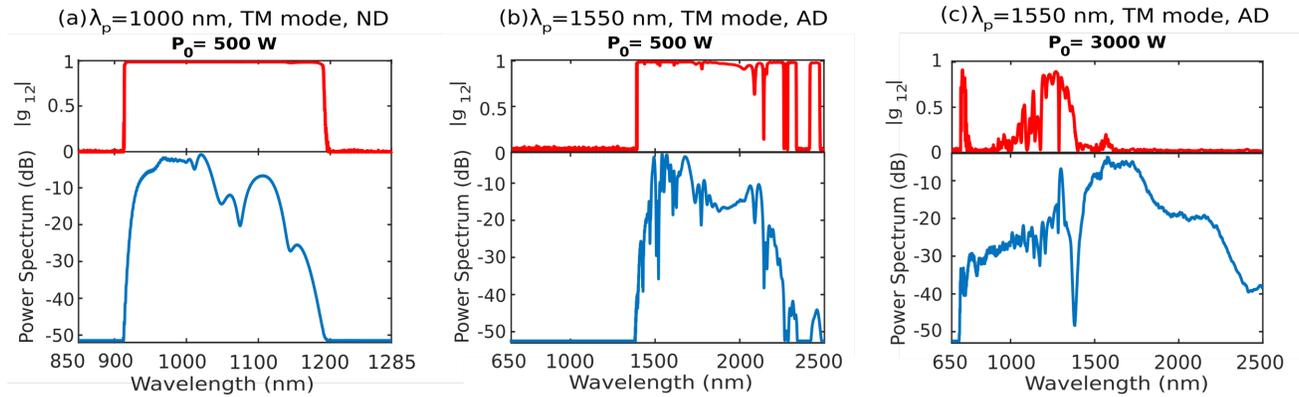

Figure 12: Numerical simulations of SC generation (blue) and the mutual coherence (red) at 1000 nm and 1550 nm, respectively, for the TM mode and different peak powers.

We also conducted SC stability analysis using the DFT technique, as described in Section 4. This method allows for the direct recording of real-time, pulse-to-pulse SC spectra in the time domain via DFT.[25] Specifically, to stretch the SC pulses, we added a 170-meter-long (normally dispersive) DCF at the chip's output fiber pigtail and then recorded the 80 MHz stretched pulse train using a high-speed, photodiode (50 GHz) paired with a 20 GHz real-time oscilloscope. Due to the photodiode's predominant responsivity from 1400 nm to 1750 nm, we confined the SC real-time analysis within this spectral range while pumping the QXP chip waveguide at 1550 nm. The DCF fiber dispersion and slope are $\beta_2$=124 ps$^2$/km and $\beta_3$=2.7 ps$^3$/km, respectively. This gives an equivalent spectral resolution close to 5 nm. The results of DFT are shown in Fig. 13 for 1024 pulses, where the pulse-to-pulse spectra are displayed in gray and the corresponding average DFT spectrum (red line) is compared to the OSA spectrum (blue line). As can be seen, the DFT average spectrum fits well with the OSA measurement for a dynamic range exceeding 20 dB.

Additionally, the DFT technique enables the straightforward measurement of spectral RIN, a crucial parameter for SC applications. Unlike the first-order spectral coherence function, the RIN formula only considers the SC intensity fluctuations without taking into account the effects of phase fluctuations. However, the RIN is a relevant metric to assess the overall SC stability, as it is described as the ratio between the standard deviation of the SC ensemble divided by its mean, as given in the following equation:[31]

$$\mathrm{RIN}(\omega) = \frac{\sqrt{\left\langle \left( \left|\tilde{A}_i(\omega)\right|^2 - \left\langle \left|\tilde{A}_i(\omega)\right|^2 \right\rangle \right)^2 \right\rangle}}{\left\langle \left|\tilde{A}_i(\omega)\right|^2 \right\rangle}. \tag{3}$$

In our case, the RIN was measured from the pulse-to-pulse spectral fluctuations obtained via DFT and further reported, in yellow, in Fig. 13. Over the SC spectral range spanning from 1500 nm to 1700 nm, we found an average RIN as low as 7 %. The results also show that the RIN increases drastically on the SC edges due to both the reduction of the SC power and the strong impact of laser input peak power fluctuations on the resulting output SC bandwidth.

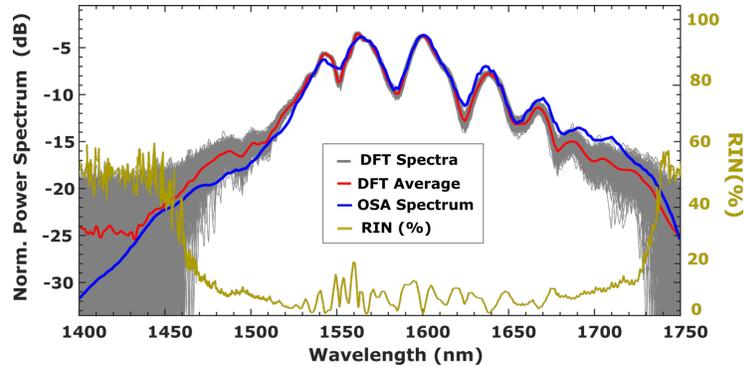

Figure 13: Real-time pulse-to-pulse spectral measurements using DFT over 1024 pulses for a pump wavelength at 1550 nm and an input power of 200 mW. DFT pulse-to-pulse spectra (light gray), DFT averaged spectrum (red), OSA spectrum (blue), and RIN retrieved from DFT measurements (RIN, yellow).

## 10. CONCLUSION

In conclusion, our study focused on supercontinuum generation in highly doped silica glass spiral waveguides of different lengths (3 cm and 50 cm, respectively) under different dispersion regimes and polarization angles. Notably, the broadest SC spectrum, spanning from 700 nm to 2400 nm, was achieved when pumped at 1550 nm. We observed that the SC generation at this wavelength was highly sensitive to the input polarization angle, whereas this sensitivity diminished at shorter pump wavelengths. Spontaneous Raman measurements of the high-index doped silica waveguides were conducted to enhance numerical modeling accuracy. Additionally, SC coherence and noise analysis demonstrated low relative intensity noise across the SC bandwidth.

These QXP glass-based integrated photonic waveguides, with their nonlinear optical properties, straightfor- ward manufacturing process, and compatibility with fiber optic technologies, show great promises for developing low-footprint and low-pulse-energy supercontinuum sources using compact femtosecond fiber lasers. Such on-chip SC sources could potentially be applied in various fields requiring broad spectral coverage, including spectroscopy, telecommunications, and biomedical imaging, among others.

## 11. ACKNOWLEDGMENTS

This work has received funding from the European Research Council (ERC) under the European Union's Horizon 2020 research and innovation program under grant agreement No. 950618 (STREAMLINE project), from the French Agence Nationale de la Recherche (ANR) through the OPTIMAL project (ANR-20-CE30-0004), and from the Conseil Régional Nouvelle-Aquitaine (SCIR & SPINAL projects). RM would like to acknowledge the NSERC Discovery and Canada Research Chair programs.

## 12. DISCLOSURES

The authors have no conflicts of interest to declare.